# Second-harmonic generation holography with polarization multiplexing for label-free collagen characterization and imaging

*Serena Goldmann,* [†] *Marie Fondanèche,* [†] *Valentina Krachmalnicoff,* [†] *Jean-Marie Chassot,* [†] *Samuel Grésillon,* [†] *Dangyuan Lei,* [◊] *Gilles Tessier,* [*,‡] *and Yannick De Wilde* [*,†]

[†] Institut Langevin, ESPCI Paris, Université PSL, 75005 Paris, France

[‡] Institut de la Vision, Sorbonne Université, 75012 Paris, France

[◊] Department of Materials Science and Engineering, Department of Physics, Centre for Functional Photonics, Hong Kong Branch of National Precious Metals Material Engineering Research Centre, and Hong Kong Institute of Clean Energy, City University of Hong Kong, 83 Tat Chee Avenue, Kowloon, Hong Kong SAR999077, China

E-mail: gilles.tessier@sorbonne-universite.fr yannick.dewilde@espci.fr

ABSTRACT

Digital holography is an interference-based imaging technique capable of recording both the amplitude and phase of an electromagnetic field. It can be obtained at the laser illumination wavelength, but also with second-harmonic generation (SHG), since the latter is produced in a coherent process. Here, we describe the development of a harmonic holographic microscope for 3D single-shot mapping of second-harmonic emitters. The knowledge of the scattered field (amplitude and phase) in a given plane (that of the camera) allows its reconstruction in any other plane using e.g. the angular spectrum representation of the optical fields, a process called 3D numerical back-propagation. In order to probe the polarization dependence of the sample's nonlinear response, we implement polarization multiplexing, in which a Wollaston prism creates two off-axis reference beams with orthogonal polarizations and non-parallel propagation directions. Each reference only interferes with the corresponding polarization component in the sample SHG emission, thus providing two independent sets of interference fringes which are easily separated in the angular spectrum representation. From a single measurement, two second-harmonic fields corresponding to orthogonal polarizations can be back-propagated. In the particular case of collagen, the second-harmonic polarization state can reveal the orientation - or disorder - of molecules and fibers. We demonstrate the feasibility of the method by reconstructing the spatial distribution of the second-harmonic field generated by collagen fibers in a rat-tail tendon sample and show that polarization-multiplexed holography can provide single-shot 3D mapping of biophysical parameters such as the helical pitch angle of collagen molecules.

INTRODUCTION

Collagen is the main structural protein in the connective tissues and is present in the extracellular matrix of most animals. As such, collagen has a major importance in fields ranging from biology, and particularly mechanobiology, to human, and even clinical applications. Collagen organization drives the biophysical and biomechanical properties of living tissues[1–6]. Specific characteristics, such as rigidity or transparency are imposed by its distribution and arrangement[7–9], which can undergo structural reorganization under mechanical stress[7] or as tissues age[10]. Many diseases are also associated with defects or changes in collagen structure. Collagen distribution and orientation can indicate, for instance, the presence or advancement state of collagen-altering conditions such as fibrosis[11,12]. Imaging collagen, in vivo or in vitro, particularly in 3-dimensions (3D) at the micrometer scale, can reveal abnormal structures and potentially serve as a diagnostic marker[4]. Like most biomolecules, collagen can be tagged using fluorescent markers, but non-invasive, non-toxic, non-photobleaching label-free approaches are valued whenever they are available, particularly for clinical applications. Among label-free optical imaging methods, optical coherence tomography (OCT) is highly efficient in terms of speed and resolution for label-free 3D imaging of biological tissues, particularly in its Full-Field (FF-OCT) implementation[13]. However, optical backscattering can be generated by a broad range of biological structures, and OCT is not specific to collagen[8,14,15] unless combined with e.g. polarization-sensitive studies[16].

Second-harmonic generation (SHG) is a nonlinear optical process which is only possible in non-centrosymmetric, non-amorphous molecules and structures. Collagen is one of the few biomolecules that generates SHG with the greatest efficiency, making SHG ideal for its specific label-free detection against a generally dark background. Indeed, since collagen proteins form a triple helix with well-aligned peptide bonds whose non-linear responses add up coherently, they exhibit a significant second order response[17–19]. SHG from biological samples is typically dominated by emission from collagen fibrils, the spatial distribution of which can be imaged using SHG microscopy[4,8,20,21]. Even in collagen, SHG is a relatively inefficient process, but since it varies quadratically with irradiance, it benefits from the use of focused pulsed laser illumination[22,23]. SHG microscopes therefore mostly operate in confocal geometry, and 3D imaging is then obtained by scanning the confocal spot over the entire sample volume. Although SHG itself does not directly involve absorption processes and heat generation (unlike e.g. fluorescence), the acceptable illumination irradiance can be limited by absorption and damage in the surrounding tissues. While SHG confocal microscopy has been demonstrated to be a powerful method to unveil the collagen structure in biological tissues[11], it bears some limitations in terms of acquisition speed, like any sequential scan-based method[24].

SHG is a coherent process. As such, it allows the use of interferometric techniques to measure both the amplitude and the phase of the generated waves. The knowledge of this electromagnetic (EM) field in a given plane allows its reconstruction in any other plane using e.g. the angular spectrum representation of optical fields[25], a process called 3D numerical back-propagation[26–30]. Among interferometric techniques, holography is arguably the most widespread. While digital holography is commonly used in the linear regime, at the illumination wavelength, the coherence of SHG signals ensures that it can also be applied to acquire SHG fields[31–41]. In holography, the field $E_O$ which interacted with the object interferes with the reference $E_R$, and the detector (usually a camera) measures an intensity $|E_O|^2 + |E_R|^2 + E_O E_R^* + E_O^* E_R$. Since $E_R$, usually a plane wave, can be known, either of the last two terms can give access to the field $E_O$. The fact that the reference $E_R$ can be intense is an important advantage when the signal $E_O$ is weak, which is often the case in SHG: the cross-product terms $E_O E_R^*$ and $E_O^* E_R$ increase with the amplitude of the reference field.

Thus, the object field benefits from an amplification effect[42], making the method particularly well-suited to measure weak SHG signals. Another important advantage of holography is that, in both the linear and nonlinear cases, it allows the entire complex field to be measured, thus giving access to the phase information, which is otherwise lost with SHG confocal microscopy[26,40], thus allowing back-propagation and 3D reconstructions of the entire EM fields from a single-shot acquisition[28,29].

The second order polarization induced by an incident field $E_\omega$ is given by $P^{(2)} \propto \chi^{(2)} E_\omega E_\omega$, where $\chi^{(2)}$ is the susceptibility tensor of the illuminated material. As such, the polarization of the SHG electric field depends on the orientation of the emitting dipoles with respect to the excitation polarization state, and can reveal the structure, orientation and degree of ordering of molecules with non-zero non-linear susceptibility, like collagen[43,44]. This provides a powerful tool to determine molecular-scale properties, e.g. the helical pitch angle of collagen molecules[44]. The most common technique to study SHG polarimetric responses is polarization-resolved SHG microscopy (P-SHG microscopy), which is intrinsically confocal in most cases, since SHG is only produced in the focal volume. P-SHG interrogates chosen elements of the $\chi^{(2)}$ tensor by rotating the polarization of the incoming fundamental beam and/or the measured polarization component[43,44]. However, acquiring (i) weak SHG signals at (ii) each position of a 2D or 3D raster-scanned image and (iii) for each polarization orientation is a relatively long process[20].

In this paper, we propose to address each of these issues by using holography to (i) amplify coherent signals and (ii) reconstruct 3D fields using (iii) two reference waves with orthogonal polarizations to simultaneously acquire holograms related to each. We demonstrate that this configuration can provide single-shot, sensitive, 3D polarization-resolved SHG images. We validate our approach on collagen-rich animal tissues (rat-tail tendon and chicken skin), in which the spatial distribution of the polarimetric response, and particularly the ratio of two non-zero components of the $\chi^{(2)}$ tensor, are used to deduce the local molecular orientation.

MULTIPLEXED POLARIZATION-RESOLVED SHG HOLOGRAPHY

SECOND HARMONIC HOLOGRAPHY SYSTEM WITH A DOUBLE REFERENCE

To investigate SHG emission, we developed a second harmonic generation (SHG) holographic microscope with multiplexed polarization analysis. The setup described in Figure 1a (and detailed in the Supplementary Information) is designed as an off-axis Mach Zehnder interferometer combined with a Ti:Sapphire laser generating 120 fs pulses at a rate of 76 MHz and wavelength λ=800 nm, divided into object and reference arms by a 50/50 nonpolarizing beam splitter. In the object arm, the fs laser beam illuminates a 23 μm diameter region of a sample containing collagen proteins placed on a glass microscope slide. The resulting SHG field is collected in transmission by a microscope objective (Numerical Aperture=0.8) and guided towards a 1024 x 1024 pixels EMCCD detector after filtering the laser excitation at the fundamental frequency. In order to allow interference on the detector with the field from the reference arm, a frequency-doubling Beta Barium Borate (BBO) crystal is introduced in the reference arm and a fine adjustment of the optical path difference between the object and reference arms within the coherence length of the fs laser is performed by means of a delay line[33]. Compared to holography with a CW laser[28,29], this adjustment is quite critical since the use of femtosecond laser pulses results in a much shorter coherence length of $35\mu m$.

Figure 1 shows how two orthogonally polarized reference beams are produced to simultaneously measure two orthogonal polarizations of the EM SHG field from the sample, taking advantage of interferometric detection. Remarkably, this approach requires no polarizing optical components in the interferometer object arm, as is the case with P-SHG microscopy. Polarization multiplexing is allowed by the simultaneous use of two reference beams carrying orthogonal polarizations and sent at different angles to interfere with the object beam. This is achieved by means of a Wollaston prism (separation angle 1°) placed in the reference arm of the interferometer. As sketched in Figure 1a, this prism splits the reference into two off-axis beams forming different angles θ$_x$ and θ$_y$ with respect to the object beam and carrying orthogonal polarizations. The object beam and the two off-axis reference beams produced in this way are contained in two intersecting planes as shown in the inset of Figure 1a. For clarity, we assume here that these planes are orthogonal, but perfect orthogonality is not required in practice.

In the plane of the EMCCD detector, defined as (O, x, y), i.e. z=0, the field $E_O$ collected by the microscope objective interferes with the two orthogonally polarized reference beams. The corresponding reference wave vectors projected in this plane can be written $k_{Rx,y} = -k_R \sin(\theta_{x,y})$. The measured intensity $I_H$, is therefore written, for each reference direction (x, y):

$$I_H = I_O + I_{Rx} + I_{Ry} + E_O^* E_{Rx} e^{-ikx.sin(\theta_x)} + E_O E_{Rx}^* e^{ikx.sin(\theta_x)} \qquad (1)$$
$$+ E_O^* E_{Ry} e^{-iky.sin(\theta_y)} + E_O E_{Ry}^* e^{iky.sin(\theta_y)}$$

Here, $I_O$ and $I_{Rx}$, $I_{Ry}$ are respectively the intensities in the object and in the reference arms, polarized along x and y. $E_O$ is the object complex amplitude (with I$_O$=E$_O$.E$_O$*), $E_{Rx}$ and $E_{Ry}$ the amplitudes of the reference waves polarized along the x and y directions, and * indicates the complex conjugate. The interference of the object field $E_O$ with the two reference beams therefore yields two interference patterns, or holograms, carried by perpendicular sets of interference fringes, along the y direction for the $E_x$ polarization and, conversely, along x for $E_y$, as illustrated by the chequerboard patterns in Figure 1a,b.

From the two superimposed holograms measured on the EMCCD detector, back propagation algorithms can be used to calculate two 3D reconstructions of the SHG field generated by the sample, each corresponding to a given polarization, $E_x$ and $E_y$. Indeed, assuming that light propagates in a homogeneous medium, the knowledge of the field (amplitude and phase) in a given plane (that of the EMCCD camera) enables its reconstruction in any other plane. Here, we use the angular spectrum representation of optical fields to calculate the 3D reconstructions[25,26,28,29]. A spatial Fourier transform is applied to the hologram, which is then expressed in the spatial frequency domain $(u_x, u_y)$. With a tilted reference beam, the Fourier spectrum of the holographic field is given by[45,46].

$$\begin{aligned}\tilde{I}_H(u_x, u_y, 0) \propto &\tilde{I}_O(u_x, u_y, 0) + \tilde{I}_R(u_x, u_y, 0) \\ &+ E_{Rx}^* \tilde{E}_{Ox}(u_x, u_y, 0) * \delta(u_x + u_{Rx}, u_y) + E_{Rx} \tilde{E}_{Ox}^*(u_x, u_y, 0) * \delta(u_x - u_{Rx}, u_y) \\ &+ E_{Ry}^* \tilde{E}_{Ox}(u_x, u_y, 0) * \delta(u_x, u_y + u_{Ry}) + E_{Ry} \tilde{E}_{Oy}^*(u_x, u_y, 0) * \delta(u_x, u_y - u_{Ry})\end{aligned} \quad (2)$$

Here, ~ indicates the Fourier transform, $\delta$ the 2D Dirac function, $*$ the convolution product and $u_{Rx,y} = k_{Rx,y}/2\pi$. Equation (2) shows that the different orders contained in the interference pattern are separated because of the angle introduced by the off-axis configuration. The information associated to the $E_x$ polarization (hologram fringes oriented parallel to the y axis) is therefore contained in diffraction orders $(+1)$ and $(-1)$, which are shifted from the center by respectively $(+u_{Rx})$ and $(-u_{Rx})$ in the spatial frequency domain. Similarly, the $E_y$-polarized SHG is shifted by $(+u_{Ry})$ and $(-u_{Ry})$. Numerically, these contributions can be easily isolated[28,29,47] and processed independently by centering each of the +1 orders, applying a chosen z-propagation term and then an inverse Fourier-transform to retrieve the amplitude and phase of either the $E_{Ox}$ or $E_{Oy}$ components, in a chosen plane of the object space.

Figure 1b shows the spatial Fourier transform of a hologram obtained by interference of the SHG field produced by a rat-tail tendon sample with two reference beams carrying orthogonal polarizations propagating along different directions after the Wollaston prism. Four bright regions delineated by violet and blue circles and corresponding to the +1 and -1 orders of each of the two linear polarizations are clearly visible. Each order can be back-propagated to multiple planes to obtain 3D reconstructions of the field amplitude, as shown in Figure 1c. In the illuminated portion of the rat-tail tendon, different structures can clearly be observed, displaying partially complementary regions, i.e. zones in which SHG is preferentially generated along the x or the y polarization direction. This can be attributed to the local density of collagen, which essentially determines the total SHG yield, and to the local orientation of its fibers, which drives the relative weights of the $E_x$ and $E_y$ components of the SHG generated in each voxel.

Remarkably, these two images corresponding to two orthogonal linear polarization states defined by the Wollaston prism result from the measurement of a single hologram. Associated with a description of the $\chi^{(2)}$ tensor of collagen, polarization-multiplexed holography can therefore enable the single-shot characterization of the orientation of collagen fibers.

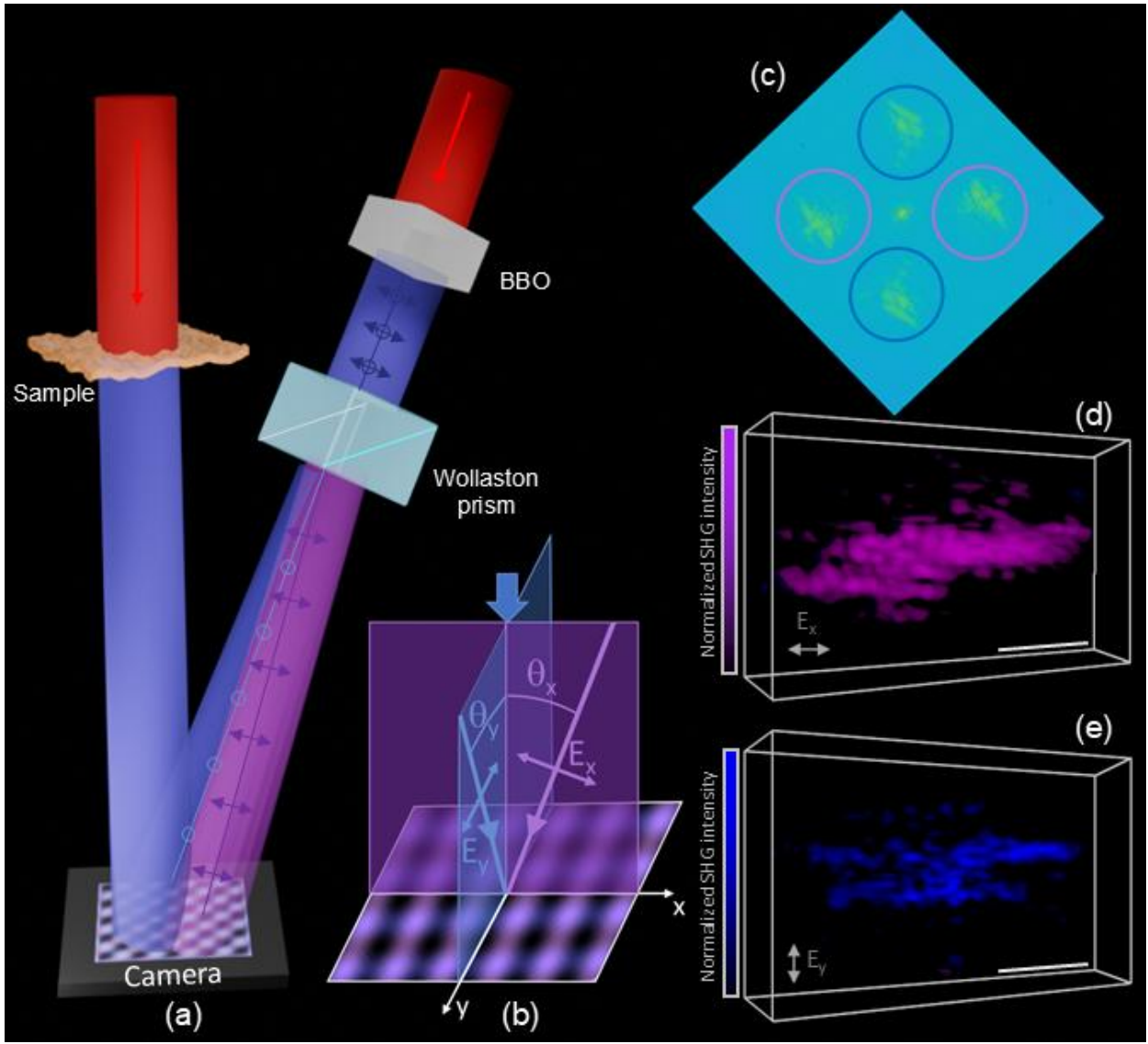

**Figure 1:** (a) Schematics of polarization multiplexing principle (BBO : Beta Barium Borate crystal). (b) Schematic showing the object beam at normal incidence on the xy plane of the EMCCD detector and the two off-axis reference beams carrying orthogonal polarizations in two secant planes. The filters eliminating the ω (red) components in the sample and reference arms have been omitted for clarity. (c) Fourier Transform of a hologram. (d, e) Independently normalized dual 3D reconstructions from a single hologram of the SHG emission generated by a rat-tail tendon aligned along the x axis, for polarizations along the X-axis (d) and Y-axis (e); scale bar: $5\mu m$.

EXPERIMENTAL CHARACTERIZATION OF COLLAGEN

This SHG holographic microscope with multiplexed polarization analysis has been used to perform a detailed study of the polarimetric response of the SHG field produced by collagen proteins in biological tissues. In this section, we first describe the link between the polarimetric response of collagen, i.e. the intensity variation of the SHG intensity for a given polarization at different angles of the polarization of the pulsed laser excitation, and the anisotropy parameter $\rho$ related to the second order nonlinear susceptibility. A detailed experimental study of the polarimetric response of collagen in rat-tail tendons for two orthogonal polarizations of the EM SHG field is then presented, from which spatial maps of collagen fiber orientation are revealed. A similar investigation applied to chicken-skin, in which collagen has a relatively disordered structure, is presented in the Supplementary Information.

DESCRIPTION OF SECOND HARMONIC GENERATION IN COLLAGEN

The second harmonic generation process is described by the second order nonlinear polarization as:

$$\vec{P}^{(2)}_{2\omega} = \epsilon_0 \chi^{(2)} \vec{E}^2_\omega \quad (3)$$

The second-order nonlinear susceptibility tensor $\chi^{(2)}$ describes the response of a non-centrosymmetric material to an incident field $E_\omega$[48]. This tensor is composed of 27 elements, but material symmetry considerations can be used to infer identical- or zero-value tensor elements.

We consider the case of type I fibrillar collagen, which has a hierarchical multiscale structure with a cylindrically symmetric arrangement[49]. In such relatively ordered tissue found in e.g. rat tail tendons, the polypeptide coils forming the collagen molecules are organized into fibrils, arranged parallel into fascicles around an axis. These fibrils are aligned along the longitudinal axis of the tendon. The fibrils themselves form 100 to 300 nm diameter cylinders aligned along the tendon axis. Here, the longitudinal axis of tendon and of the fibril cylinder is oriented along the x direction, perpendicular to the optical axis, along the z-direction. In this configuration, the $C_{6v}$ cylindrical symmetry and Kleinman's symmetry[20,44,49–52] allows to reduce drastically the number of non-zero elements in the $\chi^{(2)}$ tensor: the second order nonlinear susceptibility of collagen only contains two independent nonzero components[53], $\chi^{(2)}_{xxx}$ and $\chi^{(2)}_{xyy}\left(=\chi^{(2)}_{yxy}=\chi^{(2)}_{yyx}=\chi^{(2)}_{xzz}=\chi^{(2)}_{zxz}=\chi^{(2)}_{zzx}\right)$. Their ratio defines the anisotropy parameter $\rho$ as[19,44,50,54].

$$\rho = \frac{\chi^{(2)}_{xxx}}{\chi^{(2)}_{xyy}} \qquad (4)$$

Assuming that the fibrils are perfectly oriented along x, and that inter-fibrillar orientation disorder is negligible, the macroscopic second-order nonlinear susceptibility tensor $\chi^{(2)}$ is the sum of the individual fibril tensors $\chi^{(2)}_{fib}$ of the fibrils. Similarly, the fibril tensor is the sum of the molecule triple helix tensors $\chi^{(2)}_{TH}$ that constitute a fibril[55]. The hierarchical multiscale structure of well-aligned collagen as found in e.g. rat tail tendon consequently results in an equivalence of the anisotropy parameter at all scales[50,56,57]:

$$\rho = \frac{\chi^{(2)}_{xxx}}{\chi^{(2)}_{xyy}} = \frac{\chi^{(2)}_{xxxfib}}{\chi^{(2)}_{xyyfib}} = \frac{\chi^{(2)}_{xxxTH}}{\chi^{(2)}_{xyyTH}} \qquad (5)$$

Since the SHG emitters are the peptide bonds, their position relatively to the molecule axis defines the SHG dipole orientation, which can be characterized by a single parameter, the helical pitch angle of the collagen molecule $\theta_e$[57], and one can write:

$$\rho = \frac{\chi^{(2)}_{xxxTH}}{\chi^{(2)}_{xyyTH}} = \frac{2}{\tan^2 \theta_e} \qquad (6)$$

This expression allows to deduce molecular-scale information from the parameter, $\rho$, which can be readily measured at the macroscale.

Measuring the anisotropy parameter $\rho$ is thus an efficient way to characterize collagen structures using the SHG response of tissues[53]. The intensities of the SHG polarized along the x and y axes generated by a linearly polarized excitation along a direction $\alpha$ with respect to the collagen fibrils, thus with respect to x in our case, can be written[4,50].

$$I_{2\omega}^X \propto |\rho\cos^2\alpha + \sin^2\alpha|^2 \qquad (7)$$
$$I_{2\omega}^Y \propto |\sin 2\alpha|^2 \qquad (8)$$

Note that the orientation of the fibril can have local variations, thus locally changing the value of $\alpha$. In this case, as shown below, polarimetric SHG measurements can therefore provide a local measurement of this orientation.

However, collagen displays a strong optical anisotropy. In order to take into account its birefringence, polarization cross-talk and diattenuation, Gusachenko et al. proposed a phenomenological model which writes the total SHG intensity as[50,53]:

$$I_{2\omega} \propto \left( \left| \rho e^{-\frac{z}{\Delta l_a}} \cos^2\alpha \, e^{i\Delta\Phi} + \sin^2\alpha \right|^2 + \eta e^{-\frac{z}{\Delta l_a}} |\sin 2\alpha|^2 \right) \qquad (9)$$

with z the thickness of birefringent collagen (or other) placed in front of the probed region. Birefringence is taken into account through the phase shift $\Delta\phi = \frac{4\pi(n_e - n_o)z}{\lambda}$, and the parameter $\eta$ describes the SHG contribution due to polarization cross-talk. Finally, we define $\frac{1}{\Delta l_a} = \frac{1}{l_{\alpha=0}} - \frac{1}{l_{\alpha=\frac{\pi}{2}}}$ with $l_{\alpha=0}$ and $l_{\alpha=\frac{\pi}{2}}$ the attenuation lengths for the incident polarization components parallel and perpendicular to the sample axis. Typical values for the optical properties of collagen ($\eta, \Delta l_a, n_o, n_e$) can be found in the literature, and a measurement of the angular dependence $I_{2\omega}(\alpha)$ can therefore be used to obtain the anisotropy parameter $\rho$. In tissues containing well-aligned collagen (such as rat tail tendons), this provides a measurement of the helical pitch angle $\theta_e$. In tissues where collagen is more disordered, equations (5) and (6) do not hold, and $\rho$ merely provides a measurement of the disorder.

## INVESTIGATING THE COLLAGEN CONFORMATION IN A RAT-TAIL TENDON THROUGH THE POLARIMETRIC RESPONSE OF THE EM SHG FIELD

To demonstrate the potential of our method for tissue imaging, we study the polarimetric response of the EM SHG field produced by a rat-tail tendon, a collagen-rich tissue in which collagen fibrils[53] are aligned along a preferred orientation[55,58]. The rat-tail tendon was preserved in a phosphate-buffered saline solution and then fixed in a formaldehyde solution. Here, the incident fundamental beam at $\lambda = 800$ nm is linearly polarized along a direction forming an angle α with the horizontal x-axis. As described above, the SHG emission produced by the sample interferes with two SHG beams carrying orthogonal linear polarizations, producing two superimposed holograms on the EMCCD detector. Two 3D reconstructions of the SHG were conducted within a $25 \times 25 \times 16 \ \mu m^3$ volume, along two perpendicular linear polarizations. The reconstructed SHG intensities in Figure 1d,e clearly show the general orientation of collagen fibers along the x direction, matching the main axis of the tendon, and with an average SHG intensity 3.8 times stronger for a polarization along the axis of the tendon (x-axis) than along the perpendicular y-direction. Complementary regions of the sample can be observed, i.e. regions which preferentially generate SHG for a polarization along one polarization or the other, depending on the local orientation of collagen.

Similar measurements performed on relatively disordered tissue such as chicken skin, in which collagen is mostly randomly oriented[58], do not display this type of preferential orientation. The 3D reconstructions of the SHG measured simultaneously along two perpendicular linear polarizations on chicken skin are shown in Supplementary Information Figure S2. On average, the SHG intensity for a polarization along the x direction is 2.1 times more intense than along the y direction in chicken skin, an effect which may be linked to the fact that, in an isotropic medium, the measured intensity is expected to be greater when the excitation polarization is colinear to that of the reference. Overall, this lower value of the ratio of the x- and y-components of SHG is consistent with the fact that collagen fibrils are less organized in skin than in tendons[54,58].

In order to quantify the angular dependency of the SHG produced by collagen in a biological tissue, the next section focuses on a full polarimetric study of a rat tail tendon, which clearly displays a high degree of directionality along an easily identifiable direction.

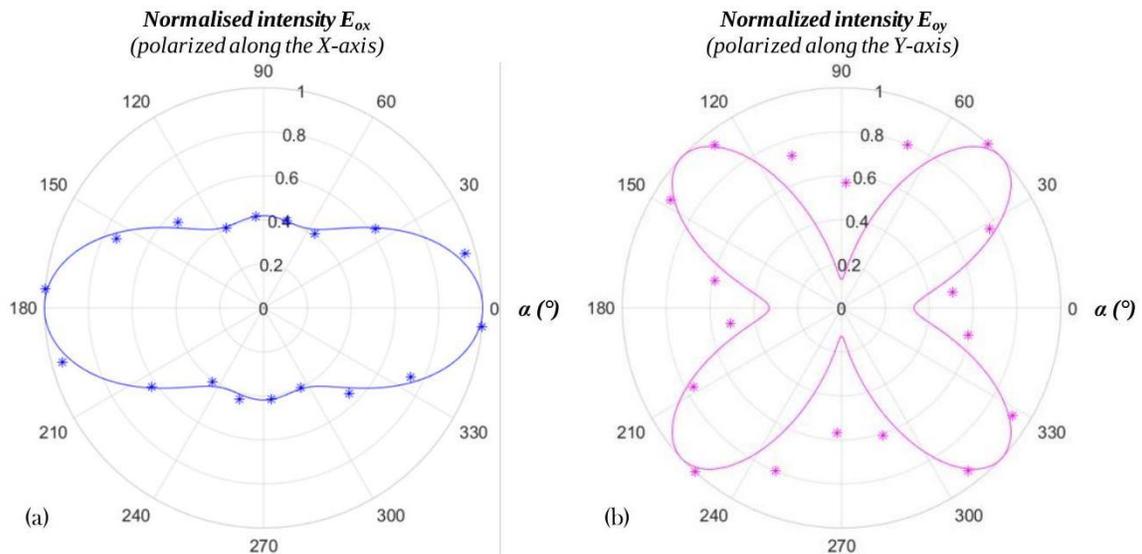

**Figure 2:** Second-harmonic generation polarized along the main axis (X, left, blue dots) of a rat-tail tendon and perpendicular to it (Y, right, pink dots) plotted against the angle $\alpha$ between the X axis and the linear-polarization of the illumination. Values are normalized for clarity, but the actual SHG emission polarized along X is 3.8 times more intense than that along Y on average. The best

fit (lines) based on equation $9^{50}$ yields $\rho = 1.6$, using parameters $z = 6\mu m, \eta_{XY} = 0.13, \Delta n = 0.0066, \Delta l_a = 175\mu m$.

Studying the second-harmonic polarization, i.e. 2 orthogonal linear components of $\vec{E}_{2\omega}$ for various orientations of the fundamental illuminating polarization is an efficient way to determine the orientation of the harmonophores, as well as their degree of organization[59]. This is achieved by measuring two orthogonal components of the SHG field while varying the angle of the illumination. By using two orthogonally polarized references in the SHG holographic microscope as sketched in Figure 1a, the polarization diagrams of the two corresponding polarization components of the SHG emission intensity from the object, $E_{Ox}$ and $E_{Oy}$, can be measured simultaneously, thus reducing by a factor of two the number of acquisitions, while benefiting from the (3D) spatial resolution of holography. In these experiments, the incident linear polarization at the fundamental frequency was rotated by an angle α with respect to the x-axis corresponding to the axis of the tendon, using a half-wave plate rotated by an angle α/2. For each position of the half-wave plate, the multiplexed interference pattern consisting of two superimposed holograms were recorded, and independent reconstructions of the 2 polarizations were performed. In every voxel of these 3D reconstructions, one can therefore record the angular dependencies of the SHG polarized along the x and y directions, $\vec{E}_{x,2\omega}(\alpha)$ and $\vec{E}_{y,2\omega}(\alpha)$ as a function of the angle α ∈[0, π]. In order to increase the signal-to-noise ratio, temporal averaging over 10 consecutive holograms was performed, in addition to a local spatial averaging within macropixels corresponding to $4.2 \times 4.2 \ \mu m^2$ in the sample space. Polar diagrams obtained on the central macropixel (2nd line, 3rd column in Figure 3 are shown in Figure 2a,b which show respectively polar plots of $\vec{E}^2_{x,2\omega}(\alpha)$ and $\vec{E}^2_{y,2\omega}(\alpha)$. Similar diagrams can be derived in each macropixel within the reconstructed region.

As expected, $\vec{E}^2_{x,2\omega}(\alpha)$ shows a clear orientation along the axis of the tendon, corresponding to the x direction, α=0. These SHG polarimetric responses are consistent with previous measurements reported using direct, sequential polarization filtering of the sample's SHG emission[50,60]. These measurements are well described by the model presented in the previous section, and a fit using Equation (9) yields a correlation coefficient of 0.97 (continuous line in Fig.2 a and b, using the same parameters). Importantly, this fit allows the extraction of the anisotropy parameter $\rho$, an operation which can be repeated in each voxel, thus providing a 2D (or 3D) map of the $\rho$ parameter over the reconstructed volume of the sample. In rat tail tendons, inter-fibrillar disorder is usually considered negligible[55], and the value of $\rho$ is determined by the local molecular orientation within the probed volume (see Equation 9). One can therefore derive 3D maps of the helical pitch angle $\theta_e$ of the collagen molecules from these measurements of the anisotropy parameter ρ.

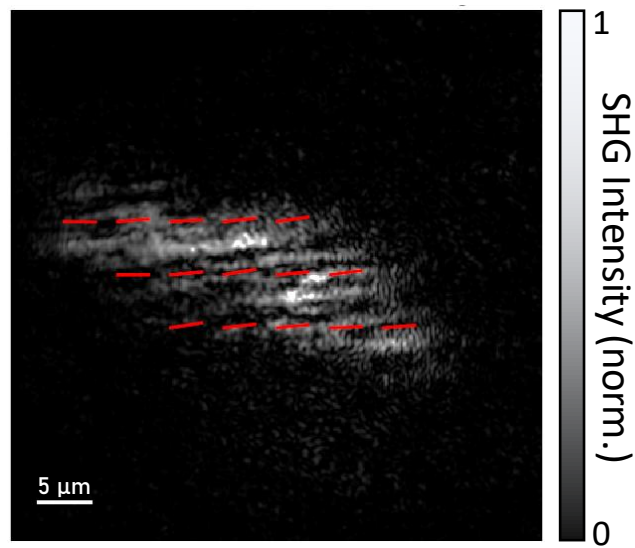

**Figure 3:** 2D cross-sectional view of a SHG image of a rat-tail tendon (grayscale), overlaid with 15 red rods oriented along the ($\theta_e$-45°) direction determined by fitting polarimetric measurements in the corresponding 4.2 × 4.2 $\mu m^2$ subregions. These rods therefore indicate the experimentally

measured deviation between the measured helical pitch angle and the expected value (45°), i.e. the local disorder of the collagen fibers (scale bar: 5 µm)

Figure 3 shows a 2D cross-section of the $\theta_e$ map deduced from our measurements from which the theoretical pitch angle value for collagen ($\theta = 45°$) has been retrieved[43]. Over this 2D cross-section, we found an average $\rho$ value of $1.49 \pm 0.14$, consistent with previous measurements, in the 1.2-2.6 range for rat tail tendons[50]. From this value of $\rho$, we deduce that the averaged helical pitch angle is 49.6°, close to the expected value of 45°. The investigated region was divided into 15 voxels, as shown in the cross-section of Figure 3, and the orientation $\theta_e$ was calculated in each sub-volume. Due to local disorder or disorientation of the collagen molecules in the tendon, this orientation displays variations within $\Delta\theta_e = 9.8°$ around the expected $\theta = 45°$ (deviation represented by the red bars in Figure 3). This value is in good agreement with previous studies of the effective orientation angle $\theta_e$ of the harmonophores in collagen rich tissues[20,43,54]. The small range of variations suggests a good structural homogeneity, which is indeed expected in rat tail tendons.

CONCLUSION

We have demonstrated a novel microscopy method which combines nonlinear holography and polarization multiplexing. By digitally processing a single holographic image obtained using two orthogonally linearly polarized reference beams in our setup, the SHG field emitted by an object can be reconstructed and decomposed into these two linear polarization components. We have applied this technique to obtain microscopic 3D images of collagen distribution in biological samples, and studied in detail the polarimetric SHG dependence of rat tail tendon collagen. Besides holographic imaging, the method has enabled one to map the SHG anisotropy parameter $\rho$ throughout the tissue, which is informative regarding the collagen structure as this parameter is sensitive to harmonophore orientation.

As holography enables full-field images to be recorded in one shot, the proposed method could be optimized to perform dynamic SHG polarimetric studies on biological tissues, for example to monitor collagen reorganization in tissue under mechanical or chemical stress. The hologram multiplexing principle we have developed can also be extended to a larger number of polarization states, towards a single-shot measurement of the Stokes vector of SHG fields.

**Funding Sources**

We acknowledge financial support from the LABEX WIFI (Laboratory of Excellence within the French Program Investments for the Future) under reference ANR-10-LABX-24 and ANR-10-IDEX-0001-02PSL* and from the Agence Nationale de la Recherche ANR-20-CE24-0021 "SP-Tunnel-OHG", as well as from the Research Grants Council of Hong Kong through an ANR/RGC JRS grant (A-CityU101/20).

ACKNOWLEDGMENT

The authors gratefully acknowledge Marie-Claire Schanne-Klein for fruitful discussions and Valérie Fradot for providing the samples.

REFERENCES

(1) Fine, S.; Hansen, W. P. Optical Second Harmonic Generation in Biological Systems. *Appl. Opt.* **1971**, *10* (10), 2350–2353. https://doi.org/10.1364/AO.10.002350.

(2) Roth, S.; Freund, I. Second Harmonic Generation in Collagen. *J. Chem. Phys.* **1979**, *70* (4), 1637–1643. https://doi.org/10.1063/1.437677.

(3) Freund, I.; Deutsch, M. Second-Harmonic Microscopy of Biological Tissue. *Opt. Lett.* **1986**, *11* (2), 94–96. https://doi.org/10.1364/OL.11.000094.

(4) Williams, R. M.; Zipfel, W. R.; Webb, W. W. Interpreting Second-Harmonic Generation Images of Collagen I Fibrils. *Biophys. J.* **2005**, *88* (2), 1377–1386. https://doi.org/10.1529/biophysj.104.047308.

(5) Chen, X.; Nadiarynkh, O.; Plotnikov, S.; Campagnola, P. J. Second Harmonic Generation Microscopy for Quantitative Analysis of Collagen Fibrillar Structure. *Nat. Protoc.* **2012**, *7* (4), 654–669. https://doi.org/10.1038/nprot.2012.009.

(6) Gusachenko, I.; Tran, V.; Houssen, Y. G.; Allain, J.-M.; Schanne-Klein, M.-C. Polarization-Resolved Second-Harmonic Generation in Tendon upon Mechanical Stretching. *Biophys. J.* **2012**, *102* (9), 2220–2229. https://doi.org/10.1016/j.bpj.2012.03.068.

(7) Ducourthial, G.; Affagard, J.-S.; Schmeltz, M.; Solinas, X.; Lopez-Poncelas, M.; Bonod-Bidaud, C.; Rubio-Amador, R.; Ruggiero, F.; Allain, J.-M.; Beaurepaire, E.; Schanne-Klein, M.-C. Monitoring Dynamic Collagen Reorganization during Skin Stretching with Fast Polarization-Resolved Second Harmonic Generation Imaging. *J. Biophotonics* **2019**, *12* (5), e201800336. https://doi.org/10.1002/jbio.201800336.

(8) Latour, G.; Gusachenko, I.; Kowalczuk, L.; Lamarre, I.; Schanne-Klein, M.-C. In Vivo Structural Imaging of the Cornea by Polarization-Resolved Second Harmonic Microscopy. *Biomed. Opt. Express* **2012**, *3* (1), 1–15. https://doi.org/10.1364/BOE.3.000001.

(9) Salameh, C.; Salviat, F.; Bessot, E.; Lama, M.; Chassot, J.-M.; Moulongui, E.; Wang, Y.; Robin, M.; Bardouil, A.; Selmane, M.; Artzner, F.; Marcellan, A.; Sanchez, C.; Giraud-Guille, M.-M.; Faustini, M.; Carminati, R.; Nassif, N. Origin of Transparency in Scattering Biomimetic Collagen Materials. *Proc. Natl. Acad. Sci.* **2020**, *117* (22), 11947–11953. https://doi.org/10.1073/pnas.2001178117.

(10) Aït-Belkacem, D.; Roche, M.; Duboisset, J.; Ferrand, P.; Brasselet, S.; Guilbert, M.; Sockalingum, G. D.; Jeannesson, P. Microscopic Structural Study of Collagen Aging in Isolated Fibrils Using Polarized Second Harmonic Generation. *J. Biomed. Opt.* **2012**, *17* (8), 080506. https://doi.org/10.1117/1.JBO.17.8.080506.

(11) Strupler, M.; Pena, A.-M.; Hernest, M.; Tharaux, P.-L.; Martin, J.-L.; Beaurepaire, E.; Schanne-Klein, M.-C. Second Harmonic Imaging and Scoring of Collagen in Fibrotic Tissues. *Opt. Express* **2007**, *15* (7), 4054–4065. https://doi.org/10.1364/OE.15.004054.

(12) Strupler, M.; Hernest, M.; Fligny, C.; Martin, J.-L.; Tharaux, P.-L.; Schanne-Klein, M.-C. Second Harmonic Microscopy to Quantify Renal Interstitial Fibrosis and Arterial Remodeling. *J. Biomed. Opt.* **2008**, *13* (5), 054041. https://doi.org/10.1117/1.2981830.

(13) Dubois, A.; Grieve, K.; Moneron, G.; Lecaque, R.; Vabre, L.; Boccara, C. Ultrahigh-Resolution Full-Field Optical Coherence Tomography. *Appl. Opt.* **2004**, *43* (14), 2874–2883. https://doi.org/10.1364/AO.43.002874.

(14) Apelian, C.; Harms, F.; Thouvenin, O.; Boccara, A. C. Dynamic Full Field Optical Coherence Tomography: Subcellular Metabolic Contrast Revealed in Tissues by Interferometric Signals Temporal Analysis. *Biomed. Opt. Express* **2016**, *7* (4), 1511–1524. https://doi.org/10.1364/BOE.7.001511.

(15) Scholler, J.; Groux, K.; Goureau, O.; Sahel, J.-A.; Fink, M.; Reichman, S.; Boccara, C.; Grieve, K. Dynamic Full-Field Optical Coherence Tomography: 3D Live-Imaging of Retinal Organoids. *Light Sci. Appl.* **2020**, *9* (1), 140. https://doi.org/10.1038/s41377-020-00375-8.

(16) Tang, P.; Kirby, M. A.; Le, N.; Li, Y.; Zeinstra, N.; Lu, G. N.; Murry, C. E.; Zheng, Y.; Wang, R. K. Polarization Sensitive Optical Coherence Tomography with Single Input for Imaging Depth-Resolved Collagen Organizations. *Light Sci. Appl.* **2021**, *10* (1), 237. https://doi.org/10.1038/s41377-021-00679-3.

(17) Campagnola, P. Second Harmonic Generation Imaging Microscopy: Applications to Diseases Diagnostics. *Anal. Chem.* **2011**, *83* (9), 3224–3231. https://doi.org/10.1021/ac1032325.

(18) Campagnola, P. J.; Millard, A. C.; Terasaki, M.; Hoppe, P. E.; Malone, C. J.; Mohler, W. A. Three-Dimensional High-Resolution Second-Harmonic Generation Imaging of Endogenous Structural Proteins in Biological Tissues. *Biophys. J.* **2002**, *82* (1), 493–508. https://doi.org/10.1016/S0006-3495(02)75414-3.

(19) Deniset-Besseau, A.; Duboisset, J.; Benichou, E.; Hache, F.; Brevet, P.-F.; Schanne-Klein, M.-C. Measurement of the Second-Order Hyperpolarizability of the Collagen Triple Helix and Determination of Its Physical Origin. *J. Phys. Chem. B* **2009**, *113* (40), 13437–13445. https://doi.org/10.1021/jp9046837.

(20) Stoller, P.; Reiser, K. M.; Celliers, P. M.; Rubenchik, A. M. Polarization-Modulated Second Harmonic Generation in Collagen. *Biophys. J.* **2002**, *82* (6), 3330–3342. https://doi.org/10.1016/S0006-3495(02)75673-7.

(21) Bancelin, S.; Couture, C.-A.; Légaré, K.; Pinsard, M.; Rivard, M.; Brown, C.; Légaré, F. Fast Interferometric Second Harmonic Generation Microscopy. *Biomed. Opt. Express* **2016**, *7* (2), 399. https://doi.org/10.1364/BOE.7.000399.

(22) Batista, A.; Breunig, H. G.; Uchugonova, A.; Morgado, A. M.; König, K. Two-Photon Spectral Fluorescence Lifetime and Second-Harmonic Generation Imaging of the Porcine Cornea with a 12-Femtosecond Laser Microscope. *J. Biomed. Opt.* **2016**, *21* (3), 036002. https://doi.org/10.1117/1.JBO.21.3.036002.

(23) Nadiarnykh, O.; Plotnikov, S.; Mohler, W. A.; Kalajzic, I.; Redford-Badwal, D.; Campagnola, P. J. Second Harmonic Generation Imaging Microscopy Studies of Osteogenesis Imperfecta. *J. Biomed. Opt.* **2007**, *12* (5), 051805. https://doi.org/10.1117/1.2799538.

(24) Gleeson, M.; Morizet, J.; Mahou, P.; Daudon, M.; Bazin, D.; Stringari, C.; Schanne-Klein, M.-C.; Beaurepaire, E. Kidney Stone Classification Using Multimodal Multiphoton Microscopy. *ACS Photonics* **2023**, *10* (10), 3594–3604. https://doi.org/10.1021/acsphotonics.3c00651.

(25) Novotny, L.; Hecht, B. *Principles of Nano-Optics*; Cambridge University Press: Cambridge, UK, 2006.

(26) Kim, M. K.; Yu, L.; Mann, C. J. Interference Techniques in Digital Holography. *J. Opt. Pure Appl. Opt.* **2006**, *8* (7), S518. https://doi.org/10.1088/1464-4258/8/7/S33.

(27) Suck, S. Y.; Collin, S.; Bardou, N.; Wilde, Y. D.; Tessier, G. Imaging the Three-Dimensional Scattering Pattern of Plasmonic Nanodisk Chains by Digital Heterodyne Holography. *Opt. Lett.* **2011**, *36* (6), 849–851. https://doi.org/10.1364/OL.36.000849.

(28) Rahbany, N.; Izeddin, I.; Krachmalnicoff, V.; Carminati, R.; Tessier, G.; De Wilde, Y. One-Shot Measurement of the Three-Dimensional Electromagnetic Field Scattered by a Subwavelength Aperture Tip Coupled to the Environment. *ACS Photonics* **2018**, *5* (4), 1539–1545. https://doi.org/10.1021/acsphotonics.7b01611.

(29) Rahbany, N.; Izeddin, I.; Krachmalnicoff, V.; Carminati, R.; Tessier, G.; De Wilde, Y. Near-Field Scanning Optical Microscope Combined with Digital Holography for Three-Dimensional Electromagnetic Field Reconstruction. In *Label-Free Super-Resolution Microscopy*; Astratov, V., Ed.; Springer International Publishing: Cham, 2019; pp 113–136. https://doi.org/10.1007/978-3-030-21722-8_5.

(30) Hu, C.; Field, J. J.; Kelkar, V.; Chiang, B.; Wernsing, K.; Toussaint, K. C.; Bartels, R. A.; Popescu, G. Harmonic Optical Tomography of Nonlinear Structures. *Nat. Photonics* **2020**, *14* (9), 564–569. https://doi.org/10.1038/s41566-020-0638-5.

(31) Yazdanfar, S.; Laiho, L. H.; So, P. T. C. Interferometric Second Harmonic Generation Microscopy. *Opt. Express* **2004**, *12* (12), 2739–2745. https://doi.org/10.1364/OPEX.12.002739.

(32) Pinsard, M.; Schmeltz, M.; Kolk, J. van der; Patten, S. A.; Ibrahim, H.; Ramunno, L.; Schanne-Klein, M.-C.; Légaré, F. Elimination of Imaging Artifacts in Second Harmonic Generation Microscopy Using Interferometry. *Biomed. Opt. Express* **2019**, *10* (8), 3938–3952. https://doi.org/10.1364/BOE.10.003938.

(33) Shaffer, E.; Pavillon, N.; Kühn, J.; Depeursinge, C. Digital Holographic Microscopy Investigation of Second Harmonic Generated at a Glass/Air Interface. *Opt. Lett.* **2009**, *34* (16), 2450–2452. https://doi.org/10.1364/OL.34.002450.

(34) Shaffer, E.; Depeursinge, C. Digital Holographic Microscopy at Fundamental and Second Harmonic Wavelengths. In *Advanced Microscopy Techniques*; SPIE, 2009; Vol. 7367, pp 122–126. https://doi.org/10.1117/12.831494.

(35) Shaffer, E.; Moratal, C.; Magistretti, P.; Marquet, P.; Depeursinge, C. Label-Free Second-Harmonic Phase Imaging of Biological Specimen by Digital Holographic Microscopy. *Opt. Lett.* **2010**, *35* (24), 4102–4104. https://doi.org/10.1364/OL.35.004102.

(36) Shaffer, E.; Marquet, P.; Depeursinge, C. Second Harmonic Phase Microscopy of Collagen Fibers. In *Multiphoton Microscopy in the Biomedical Sciences XI*; SPIE, 2011; Vol. 7903, pp 63–68. https://doi.org/10.1117/12.874538.

(37) Pu, Y.; Psaltis, D. Digital Holography of Second Harmonic Signal. In *Conference on Lasers and Electro-Optics/Quantum Electronics and Laser Science Conference and Photonic Applications Systems Technologies (2006), paper CFA5*; Optica Publishing Group, 2006; p CFA5.

(38) Pu, Y.; Psaltis, D. Seeing through Turbidity with Harmonic Holography [Invited]. *Appl. Opt.* **2013**, *52* (4), 567–578. https://doi.org/10.1364/AO.52.000567.

(39) Pu, Y.; Centurion, M.; Psaltis, D. Harmonic Holography: A New Holographic Principle. *Appl. Opt.* **2008**, *47* (4), A103–A110. https://doi.org/10.1364/AO.47.00A103.

(40) Pu, Y.; Psaltis, D. Second-Harmonic Radiating Imaging Probes and Harmonic Holography. In *Ultrafast Nonlinear Imaging and Spectroscopy IV*; SPIE, 2016; Vol. 9956, pp 20–32. https://doi.org/10.1117/12.2238511.

(41) Smith, D. R.; Winters, D. G.; Bartels, R. A. Submillisecond Second Harmonic Holographic Imaging of Biological Specimens in Three Dimensions. *Proc. Natl. Acad. Sci.* **2013**, *110* (46), 18391–18396. https://doi.org/10.1073/pnas.1306856110.

(42) Picart, P.; Gross, M.; Coëtmellec, S.; Lebrun, D.; Brunel, M.; Desse, J. M.; El Mallahi, A.; Minetti, C.; Dubois, F.; Rappaz, B.; Pavillon, N.; Georges, M.; Verrier, N.; Atlan, M.; Marquet, P. *New Techniques in Digital Holography*, ISTE-Wiley.; London, 2015.

(43) Plotnikov, S. V.; Millard, A. C.; Campagnola, P. J.; Mohler, W. A. Characterization of the Myosin-Based Source for Second-Harmonic Generation from Muscle Sarcomeres. *Biophys. J.* **2006**, *90* (2), 693–703. https://doi.org/10.1529/biophysj.105.071555.

(44) Psilodimitrakopoulos, S.; Amat-Roldan, I.; Loza-Alvarez, P.; Artigas, D. Estimating the Helical Pitch Angle of Amylopectin in Starch Using Polarization Second Harmonic Generation Microscopy. *J. Opt.* **2010**, *12* (8), 084007. https://doi.org/10.1088/2040-8978/12/8/084007.

(45) Picart, P.; Leval, J. General Theoretical Formulation of Image Formation in Digital Fresnel Holography. *JOSA A* **2008**, *25* (7), 1744–1761. https://doi.org/10.1364/JOSAA.25.001744.

(46) Verrier, N.; Alexandre, D.; Tessier, G.; Gross, M. Holographic Microscopy Reconstruction in Both Object and Image Half-Spaces with an Undistorted Three-Dimensional Grid. *Appl. Opt.* **2015**, *54* (15), 4672–4677. https://doi.org/10.1364/AO.54.004672.

(47) Cuche, E.; Marquet, P.; Depeursinge, C. Spatial Filtering for Zero-Order and Twin-Image Elimination in Digital off-Axis Holography. *Appl. Opt.* **2000**, *39* (23), 4070–4075. https://doi.org/10.1364/AO.39.004070.

(48) *Second Harmonic Generation Imaging*; Pavone, F. S., Campagnola, P. J., Eds.; CRC Press: Boca Raton, 2013. https://doi.org/10.1201/b15039.

(49) Lee, A. H.; Elliott, D. M. Comparative Multi-Scale Hierarchical Structure of the Tail, Plantaris, and Achilles Tendons in the Rat. *J. Anat.* **2019**, *234* (2), 252–262. https://doi.org/10.1111/joa.12913.

(50) Gusachenko, I.; Latour, G.; Schanne-Klein, M.-C. Polarization-Resolved Second Harmonic Microscopy in Anisotropic Thick Tissues. *Opt. Express* **2010**, *18* (18), 19339–19352. https://doi.org/10.1364/OE.18.019339.

(51) Reiser, K.; Stoller, P.; Knoesen, A. Three-Dimensional Geometry of Collagenous Tissues by Second Harmonic Polarimetry. *Sci. Rep.* **2017**, *7* (1), 2642. https://doi.org/10.1038/s41598-017-02326-7.

(52) Alizadeh, M.; Merino, D.; Lombardo, G.; Lombardo, M.; Mencucci, R.; Ghotbi, M.; Loza-Alvarez, P. Identifying Crossing Collagen Fibers in Human Corneal Tissues Using pSHG Images. *Biomed. Opt. Express* **2019**, *10* (8), 3875–3888. https://doi.org/10.1364/BOE.10.003875.

(53) Gusachenko, I.; Schanne-Klein, M.-C. Numerical Simulation of Polarization-Resolved Second-Harmonic Microscopy in Birefringent Media. *Phys. Rev. A* **2013**, *88* (5), 053811. https://doi.org/10.1103/PhysRevA.88.053811.

(54) Tiaho, F.; Recher, G.; Rouède, D. Estimation of Helical Angles of Myosin and Collagen by Second Harmonic Generation Imaging Microscopy. *Opt. Express* **2007**, *15* (19), 12286–12295. https://doi.org/10.1364/OE.15.012286.

(55) Krins, N.; Wien, F.; Schmeltz, M.; Pérez, J.; Dems, D.; Debons, N.; Laberty-Robert, C.; Schanne-Klein, M.-C.; Aimé, C. Angle-Resolved Linear Dichroism to Probe the Organization of Highly Ordered Collagen Biomaterials. *Biomacromolecules* **2024**, *25* (9), 6181–6187. https://doi.org/10.1021/acs.biomac.4c00860.

(56) Tuer, A. E.; Krouglov, S.; Prent, N.; Cisek, R.; Sandkuijl, D.; Yasufuku, K.; Wilson, B. C.; Barzda, V. Nonlinear Optical Properties of Type I Collagen Fibers Studied by Polarization Dependent Second Harmonic Generation Microscopy. *J. Phys. Chem. B* **2011**, *115* (44), 12759–12769. https://doi.org/10.1021/jp206308k.

(57) Tuer, A. E.; Akens, M. K.; Krouglov, S.; Sandkuijl, D.; Wilson, B. C.; Whyne, C. M.; Barzda, V. Hierarchical Model of Fibrillar Collagen Organization for Interpreting the Second-Order Susceptibility Tensors in Biological Tissue. *Biophys. J.* **2012**, *103* (10), 2093–2105. https://doi.org/10.1016/j.bpj.2012.10.019.

(58) Yasui, T.; Tohno, Y.; Araki, T. Characterization of Collagen Orientation in Human Dermis by Two-Dimensional Second-Harmonic-Generation Polarimetry. *J. Biomed. Opt.* **2004**, *9* (2), 259–264. https://doi.org/10.1117/1.1644116.

(59) Yasui, T.; Tohno, Y.; Araki, T. Determination of Collagen Fiber Orientation in Human Tissue by Use of Polarization Measurement of Molecular Second-Harmonic-Generation Light. *Appl. Opt.* **2004**, *43* (14), 2861–2867. https://doi.org/10.1364/AO.43.002861.

(60) Aït-Belkacem, D.; Gasecka, A.; Munhoz, F.; Brustlein, S.; Brasselet, S. Influence of Birefringence on Polarization Resolved Nonlinear Microscopy and Collagen SHG Structural Imaging. *Opt. Express* **2010**, *18* (14), 14859–14870. https://doi.org/10.1364/OE.18.014859.

# Supplementary information

DETAILED DESCRIPTION OF THE SHG OFF-AXIS HOLOGRAPHIC MICROSCOPE WITH POLARIZATION MULTIPLEXING

The SHG holographic microscope is described in figure S1. It is designed as an off-axis Mach Zehnder interferometer. A Ti:Sapphire laser generating 120 fs pulses at a rate of 76 MHz is divided into object and reference arms by a 50/50 nonpolarizing beam splitter (P = 400 mW average power each). In the object arm, the pulsed laser beam centered at $\lambda = 800$ nm is focused by a f' = 8 cm lens on the sample. This results in a relatively broad illuminated region, $23\mu$m in diameter, which is essential to full-field imaging. The frequency-doubled Electro_Magnetic (EM) field generated by the sample is collected by a X100/0.8NA microscope objective (Olympus). In the reference arm, the laser is focused on a Beta Barium Borate (BBO) doubling crystal and recollimated by a second confocal lens.

Both reference beams are expanded to match the diameter of the sample using a beam expander. Blue filters (Schott,BG40) are placed in both arms to suppress the remaining contributions of the fundamental EM field. While this filtering is carried out in both arms to avoid saturating the camera with light at $\lambda$, an efficient filtering of the fundamental beam in just one arm is in principle sufficient to cancel out any interference at $\lambda$. This is done preferentially in the reference arm (4 filters BG40 are superimposed, resulting in a total transmission of $(5.10^{-6})^4 \approx 10^{-20}$ at $\lambda$, and $0.83^4=0.48$ at $\lambda/2$) where the available optical power is sufficient to illuminate the camera. In this way, we minimize unwanted attenuation of $2\omega$ frequency (*ie* $\lambda/2$) signals in the sample arm. The filtered beams are then recombined with a second beam splitter and interfere on a 1024 x 1024 pixels EM-CCD detector (Andor) in an off - axis configuration. Compared to holography with a CW laser[1,2], the use of femtosecond laser pulses results in a much shorter coherence length of 35$\mu$m. A delay line is therefore required in the setup to find the zero-path difference and enter the coherence regime between the reference and sample beams[3].

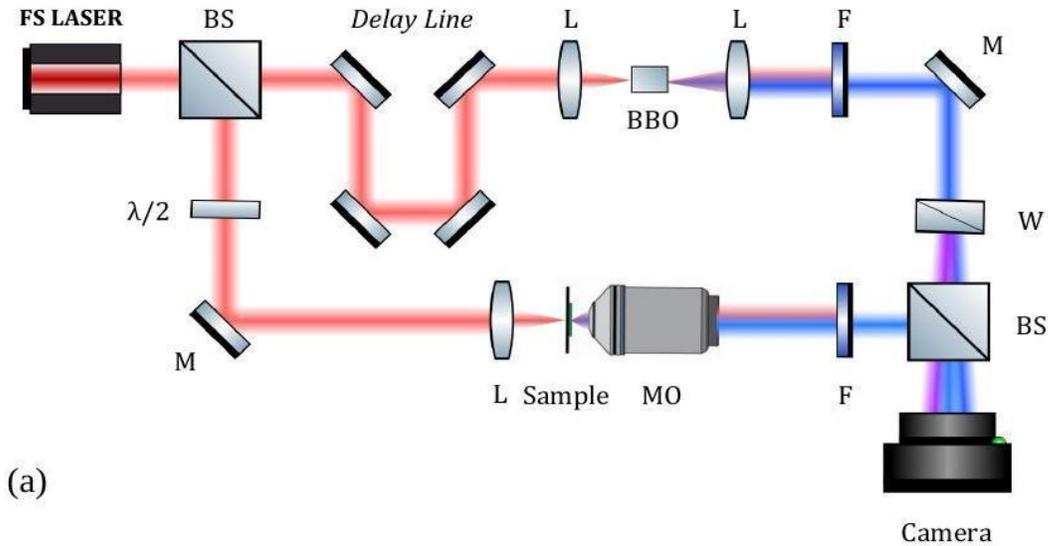

**Figure S1:** Optical set-up for dual-polarization off-axis holography; BS, beam splitter; M, mirror; L, Lens; MO, Microscope objective; F, filter; $\lambda/2$, half wave plate; BBO, Beta Barium Borate crystal; W, Wollaston prism. The delay line allows to adjust the optical path difference of the object and reference arms, to enable the interference of low-coherence-length, ultrashort laser pulses. The Wollaston prism generates two off-axis reference beams carrying orthogonal polarizations.

3D RECONSTRUCTIONS OF THE SHG SIMULTANEOUSLY MEASURED ON CHICKEN SKIN ALONG TWO ORTHOGONAL POLARIZATIONS

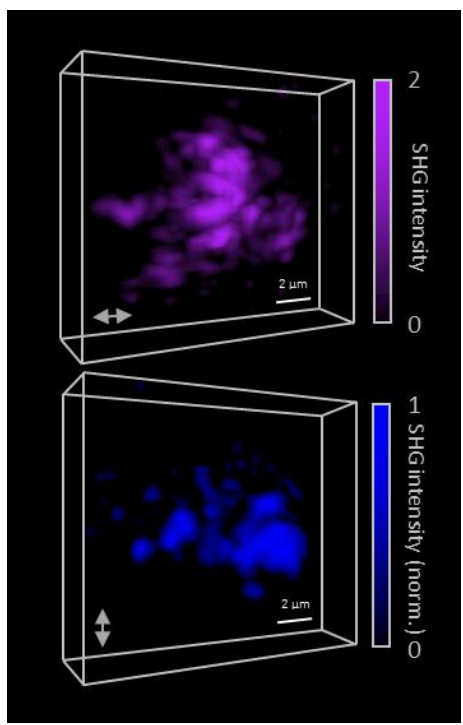

**Figure S2:** Dual 3D reconstructions from a single hologram of the SHG emission generated by chicken skin illuminated at 800 nm with a polarization along the x (horizontal) direction. The chicken skin fragment was positioned on a glass slide, in the (x, y) plane. SHG was measured along 2 perpendicular orientations (x: pink, y: blue). The colorbars are normalized to the maximum signal along the y direction(blue), with the x-polarized signal (pink) approximately 2.1 times stronger.

Chicken skin, like that of most vertebrates, mostly contains entangled collagen fibers, highly variable in their organization and orientation4. The dermis, which is surrounded by the epidermis and hypodermis layers is rich in collagen fibers which have no dominant orientation. As opposed to tendons which are strongly unidirectional, skin samples are known to exhibit little anisotropy in the x-y plane.

Second Harmonic holographic measurements have been conducted on chicken skin samples, positioned in the (x, y) plane on a glass slide and held using magnets to avoid tensions or constraints in the imaged area. This sample was illuminated along the z direction with a polarization along the x direction. 3D reconstructions of the SHG recorded simultaneously within a 25×25×16 μm^3 volume are shown in fig. S2.

Locally, complementary regions of the sample can be observed, i.e. regions which preferentially generate SHG for a polarization along one polarization or the other, depending on the local orientation of collagen. On average, the SHG intensity for a polarization along the X direction is 2.1 more intense than along the Y direction, an effect which may be linked to the fact that, in an isotropic medium, the measured intensity is expected to be greater when the excitation polarization is colinear to that of the reference.

As opposed to rat tendons (see main text), polarization diagrams recorded in chicken skin display a response close to that of a anisotropic medium, attributed to the strong local inhomogeneity in the fiber orientation at the microscopic scale. Fig. S3 shows polarization diagrams of 4 independent polarization-resolved measurements of collagen response in chicken dermis. As expected in randomly distributed collagen, there is no specific prominent orientation inferred by the polarization diagrams. However, more measurements should be conducted in order to better characterize the chicken dermis response.

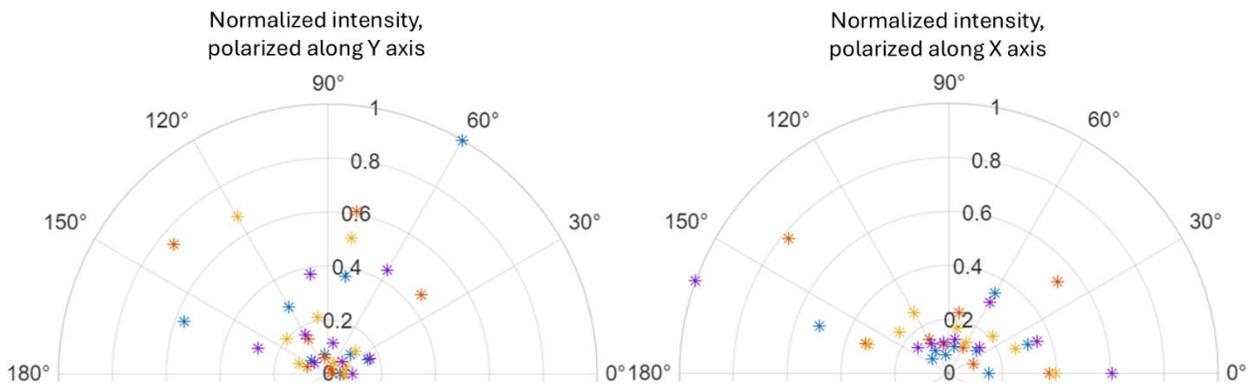

**Figure S3:** Polarization diagrams measured on chicken skin, conducted in neighboring regions.

BIBLIOGRAPHY

(1) Suck, S. Y.; Collin, S.; Bardou, N.; Wilde, Y. D.; Tessier, G. Imaging the Three-Dimensional Scattering Pattern of Plasmonic Nanodisk Chains by Digital Heterodyne Holography. *Opt. Lett.* **2011**, *36* (6), 849–851. https://doi.org/10.1364/OL.36.000849.

(2) Rahbany, N.; Izeddin, I.; Krachmalnicoff, V.; Carminati, R.; Tessier, G.; De Wilde, Y. One-Shot Measurement of the Three-Dimensional Electromagnetic Field Scattered by a Subwavelength Aperture Tip Coupled to the Environment. *ACS Photonics* **2018**, *5* (4), 1539–1545. https://doi.org/10.1021/acsphotonics.7b01611.

(3) Yu, W.; Li, X.; Hu, R.; Qu, J.; Liu, L. Full-Field Measurement of Complex Objects Illuminated by an Ultrashort Pulse Laser Using Delay-Line Sweeping off-Axis Interferometry. *Opt. Lett. Vol 46 Issue 12 Pp 2803-2806* **2021**. https://doi.org/10.1364/OL.421313.

(4) Aghigh, A.; Bancelin, S.; Rivard, M.; Pinsard, M.; Ibrahim, H.; Légaré, F. Second Harmonic Generation Microscopy: A Powerful Tool for Bio-Imaging. *Biophys. Rev.* **2023**, *15* (1), 43–70. https://doi.org/10.1007/s12551-022-01041-6.